\documentclass[twocolumn,letter]{jpsj2}

\usepackage{bm}

\title{%
Efficiency of Energy Transduction in a Molecular Chemical Engine
}

\author{%
Kazuo \textsc{Sasaki}\thanks{
E-mail address: sasaki@camp.apph.tohoku.ac.jp.}, 
Ryo \textsc{Kanada}$^1$ and Satoshi \textsc{Amari}
}

\inst{%
Department of Applied Physics, Tohoku University, 
6-6-05 Aoba-yama, Sendai 980-8579 \\
$^1$Cybermedia Center, 
Osaka University,Toyonaka 560-0043
}

\recdate{November 10, 2006}

\abst{%
A simple model of the two-state ratchet type is proposed
for molecular chemical engines that convert chemical
free energy into mechanical work and vice versa.  
The engine works by catalyzing a chemical reaction and 
turning a rotor.  Analytical expressions are obtained for the 
dependences of rotation and reaction rates on the 
concentrations of reactant and product molecules, 
from which the performance of the engine is analyzed. 
In particular, the efficiency of energy transduction 
is discussed in some detail. 
}

\kword{%
molecular chemical engine, molecular motor, ratchet, efficiency, mechanochemical coupling}

\begin{document}

\sloppy
\maketitle

The F$_1$ motor (F$_1$-ATPase), which is one of biological 
molecular motors \cite{alberts2002, schliwa2003}, 
 is a remarkable molecular machine.  
 It  works as a rotary motor   
 when it catalyzes the hydrolysis of adenosine triphosphate
 (ATP) into adenosine diphosphate (ADP) and  inorganic phosphate (Pi) 
 \cite{noji97, yasuda2001}.
It can also generate (synthesize) ATP from ADP and Pi when its rotor is
forced to rotate in the opposite direction 
\cite{itoh2004, rondelez2005}; 
this is analogous to a heat engine working as a heat pump.  
A molecular machine, like the F$_1$ motor, 
that can convert chemical energy into mechanical work 
and vice versa will be referred to as a \textit{molecular chemical engine}; 
it operates as a \textit{motor} if it produces motion out of chemical energy, 
whereas it works as a \textit{generator} if it generates ``fuel'' molecules 
of high chemical potential from ``waste'' molecules of low chemical potential  
by consuming mechanical energy. 

It has been recognized that certain fundamental features of biological 
molecular motors can be captured by ``Brownian motor'' or ratchet models
in which the system undergoes Brownian motion on a potential surface 
that changes stochastically between two or more profiles 
corresponding to different chemical states;  
see, for example, refs.~\citen{astumian97, julicher97, 
bustamante2001, reimann2002, astumian2005} for reviews. 
For example, simple two-state ratchet models have demonstrated 
how the Brownian motion can be rectified to produce directed 
motion \cite{prost94, astumian94}; 
the dependence of the motor velocity on the concentration 
of the fuel molecule ATP have been analyzed with simple models 
\cite{lipowsky2000, lipowsky2003, lattanzi2001} 
and with an elaborate model \cite{sun2004}; 
and the efficiency of energy transduction \cite{parmeggiani99} 
and other measures of efficiency \cite{derenyi99, wang2002, wang2005} 
have been discussed. 
Although the effects of the concentration of fuel molecule or 
the input free energy 
on the performance of molecular motors have been studied  
in previous investigations,  
little attention has been payed to the effects of waste molecules. 

In this letter we propose a simple model 
for molecular chemical engines that explicitly takes into 
account the effects of both the reactant and product 
molecules. 
Our model is a variant of two-state ratchet models 
\cite{astumian97, julicher97, astumian2005, prost94, 
astumian94} in general, 
and is closely related with the one outlined 
by Astumian and Bier \cite{astumian96} in particular.
The main difference between our model and 
other two-state models worked out earlier lies in the physics associated  
with transitions between the states: 
in the  present model a transition occurs when the engine binds or releases 
a fuel or waste molecule, whereas it occurs as a result of ATP hydrolysis or  
the reverse reaction 
in some versions  
\cite{parmeggiani99, 
lipowsky2000, lipowsky2003, lattanzi2001} 
of previous models; 
in another \cite{astumian96} 
the transitions represents binding and unbinding of nucleotides, 
but the way how the nucleotide concentrations are related with 
the transition rates are different from ours. 
Analytic expressions will be obtained for the dependences
of rotation and reaction rates of the engine on the 
concentrations of reactant and product molecules. 
By using these results, the efficiency of energy transduction
is studied for both the cases of the engine operating as 
a motor and as a generator.


Our model engine has a rotor, whose rotational angle is denoted by $\theta$. 
The conformation (structure) of the engine is assumed to change 
as $\theta$ is varied. 
The engine catalyzes a hypothetical chemical reaction $A \rightleftarrows B$ 
instead of more complicated reaction $\text{ATP} \rightleftarrows \text{ADP} + \text{Pi}$ 
taking place in biological molecular motors. 
The chemical potentials $\mu_A$ and $\mu_B$ of molecules $A$ and $B$, 
respectively, in the isothermal environment of temperature $T$ are assumed to satisfy 
the condition $\Delta \mu \equiv \mu_A - \mu_B > 0$; 
i.e., $A$ is the fuel and $B$ is the waste.  
The engine can bind at most one molecule $A$ or $B$. 
For simplicity, 
the binding and dissociation of  $A$ or $B$ are 
supposed to be possible only if $\theta$ 
is at a particular 
value $\theta_A$ or $\theta_B$, respectively, with $0 < \theta_A < \theta_B < 2\pi$, 
i.e., if the engine is in a particular conformation. 

We adopt the following assumption \cite{sasaki2005} on 
the reaction taking place in the engine: 
the relaxation of the ``reaction coordinate'' describing this reaction 
is so quick that this coordinate is always in thermal equilibrium, 
and the forward (backward) reaction $A \to B$ ($B \to A$) proceeds with certainty 
when $\theta$ is varied from $\theta_A$ to $\theta_B$  
($\theta_B$ to $\theta_A$). 
This will allow us to define an effective potential $V_1(\theta)$ 
 (it is actually a free energy \cite{hill74, sasaki2005}) 
of the engine when it carries a ``ligand molecule,'' 
which is  $A$ for $\theta \sim \theta_A$ and $B$ for $\theta \sim \theta_B$. 
(Alternatively, the effective potential $V_1$ may be 
reduced from more detailed models 
\cite{wang98, sun2004, xing2005} of the F$_1$ motor, 
for example.)

\begin{figure}
	\begin{center}
		\includegraphics[width = 7.0 cm]{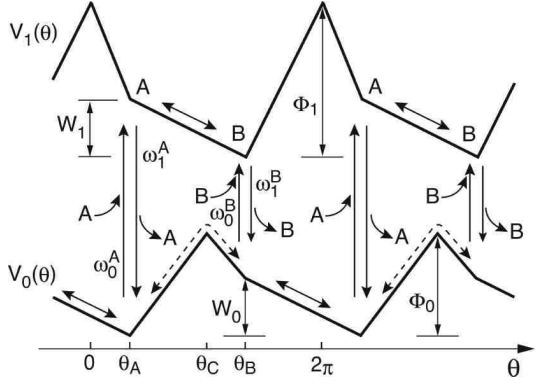}
	\end{center}
	\vspace{-2mm}
	\caption{
		Schematic representation of the present model for a molecular chemical engine: 
		an example of a pair of potentials $V_0(\theta)$ and $V_1(\theta)$ in 
		states 0 and 1, respectively, is shown together with possible pathways, 
		indicated by arrows, which the engine can take. 
		See the text for details. 
	\label{fig:potentials}}
\end{figure}

The engine is said to be in ``state 1'' or ``state 0'' depending, respectively, 
on whether it is occupied by a ligand molecule or not. 
Let $V_0(\theta)$ be the potential (free energy) of the engine in state 0. 
Then the torque  exerted on the rotor by the engine in state $j$ ($j = 0, 1$) 
is given by $-{\rm d}V_j/{\rm d}\theta$. 
In Fig.~\ref{fig:potentials}, an example of the pair of potentials $V_0$ and $V_1$ 
(which is partly motivated by the analysis \cite{kinosita2004} 
of experiments on the F$_1$ motor)
is shown, together with ``pathways'' (indicated by arrows) the engine can take. 
For simplicity, 
we assume that a \textit{large barrier} of potential $V_1$ 
lies in the interval $\theta_B < \theta < \theta_A + 2\pi$ (mod $2\pi$) 
which cannot be surmounted by the rotor; 
we also assume that potential $V_0$ has a barrier, 
which may be large or small,  
in the interval $\theta_A < \theta < \theta_B$. 
If the latter barrier is so large that the dashed passes in Fig.~\ref{fig:potentials} 
cannot be taken, 
a single forward revolution (increase in $\theta$ by $2\pi$) of the rotor 
is always accompanied by a single forward chemical reaction $A \to B$; 
if this is the case, 
it is said that the ``mechanochemical coupling'' of the engine is \textit{tight}. 
Otherwise, it is \textit{loose} \cite{oosawa86, bustamante2001}. 

A transition between states~0 and 1 occurs when 
the engine binds or releases a ligand molecule 
(vertical arrows in Fig.~\ref{fig:potentials}).  
Let $w_j(\theta)$ be the rate of transition from state $j$ 
to the other state at angle $\theta$ of the rotor. 
The assumption mentioned above 
that the binding and the dissociation occur at particular values of $\theta$ 
may be expressed as 
\begin{equation}
	w_j(\theta) = \omega_j^A\delta(\theta - \theta_A)
		+ \omega_j^B\delta(\theta - \theta_B)
	\quad (j = 0, 1), 
	\label{eq:wj}
\end{equation}
where $\omega_j^A$ and $\omega_j^B$ are positive constants 
associated with binding ($j = 0$) and dissociation ($j = 1$) of 
molecules $A$ and $B$ (see Fig.~\ref{fig:potentials}), 
and $\delta(\theta)$ is the delta function. 
The use of the delta function is a mathematical 
idealization,  
which has been introduced by several authors 
\cite{julicher97, derenyi99, bagdassarian2000, 
lipowsky2000, lipowsky2003, woo2005, woo2006} 
to carry out various calculations analytically.
	
The condition of detailed balance requires \cite{hill74} the first equality in 
\begin{equation}
	\frac{\omega_0^\alpha}{\omega_1^\alpha} 
	= \exp\left(\frac{\mu_\alpha - \Delta V_\alpha}{k_\text{B}T}\right) 
	= \frac{n_\alpha}{n_\alpha^0} = \rho_\alpha
	\quad (\alpha = A, B)
	\label{eq:dbalance}
\end{equation}
to hold, where $\alpha = A$ or $B$,  
$\Delta V_\alpha = V_1(\theta_\alpha) - V_0(\theta_\alpha)$, 
and $k_\text{B}$ is the Boltzmann constant. 
The second equality in eq.~(\ref{eq:dbalance}) comes from the fact 
that the concentration $n_\alpha$ of molecule $\alpha$ is proportional to 
$\exp(\mu_\alpha/k_\text{B}T)$ ($n_\alpha^0$ is the value of $n_\alpha$ 
for which $\mu_\alpha = \Delta V_\alpha$), 
and the third equality defines the reduced concentration $\rho_\alpha$. 
The binding rate of  molecule $\alpha$ will be assumed to 
be proportional to $\rho_\alpha$; 
this implies that the dissociation rate is independent of $\rho_\alpha$ 
due to eq.~(\ref{eq:dbalance}).  
Hence we have
\begin{equation}
	\omega_0^\alpha = \kappa_\alpha \rho_\alpha, 
	\quad \omega_1^\alpha = \kappa_\alpha
	\qquad (\alpha = A, B)
	\label{eq:kappa}
\end{equation}
with some positive constant $\kappa_\alpha$. 

We shall consider a situation in which a constant torque $L \ge 0$, 
opposing the forward rotation,  
is exerted on the rotor  externally. 
If the rotor revolves forward, 
the engine does mechanical work against the load torque 
(it works as a motor). 
On the other hand, if it revolves backward  
\textit{and} the backward reaction $B \to A$ takes place, 
the engine works as a generator.  

The rotational motion of the rotor is supposed to be 
described by the Langevin equation in the overdamped limit 
\cite{reimann2002}. 
Then, the rotation rate $\nu$ 
[the average of $({\rm d}\theta/{\rm d}t)/2\pi$] 
and the reaction rate $r$ (the difference of the average numbers 
of forward and backward reactions per unit time) may be 
calculated from the steady-state solution to the Smoluchowski 
(Fokker-Planck) equation 
\begin{equation}
	\frac{\partial P_j}{\partial t} + \frac{\partial J_j}{\partial \theta}
	= - w_jP_j + w_{1 - j}P_{1 - j} 
	\quad (j = 0, 1),
	\label{eq:fpeq}
\end{equation}
associated with the Langevin equation, 
for the probability densities $P_j(\theta, t)$ of $\theta$  in state $j$ 
at time $t$. 
In eq.~(\ref{eq:fpeq}), $w_j$ is the transition rate given in eq.~(\ref{eq:wj}) and 
$J_j = -D_0(\partial P_j/\partial\theta) - 
({\rm d}V_j/{\rm d}\theta + L)(P_j/\gamma)$ 
is the probability current density in state $j$ with 
$D_0 = k_\text{B}T/\gamma$ the diffusion coefficient associated
with the free rotation of the rotor, 
and $\gamma$ the drag coefficient of the rotor.
Provided that the steady-state solution $P_j(\theta) = P_j(\theta + 2\pi)$ is normalized 
in such a way that 
$\int_0^{2\pi}(P_0 + P_1)\,{\rm d}\theta = 1$, 
we have $\nu = J_0 + J_1$ and 
$r = \omega_0^AP_0(\theta_A) - \omega_1^AP_1(\theta_A) 
= \omega_1^BP_1(\theta_B) - \omega_0^BP_0(\theta_B)$; 
note that $J_0 + J_1$ is independent of $\theta$ in the steady state.


The steady-state solution can be obtained analytically \cite{julicher97, derenyi99, bagdassarian2000, 
lipowsky2000, lipowsky2003, woo2005, woo2006}
for the Smoluchowski equation~(\ref{eq:fpeq}) with 
transition rates $w_j(\theta)$ given as sums of the delta functions. 
In the present model, in which the rate constants are given by eq.~(\ref{eq:kappa}), 
the following results for $\nu$ and $r$ are obtained: 
\begin{align}
	\frac{\nu}{D_0} &= \frac{\rho_A - \sigma\rho_B - a}
	{c_0 + c_A\rho_A + c_B\rho_B + c_{AB}\rho_A\rho_B},
	\label{eq:v} \\
	\frac{r}{D_0} &= \frac{b_A\rho_A - \sigma b_B\rho_B}
		{c_0 + c_A\rho_A + c_B\rho_B + c_{AB}\rho_A\rho_B},
	\label{eq:r}
\end{align}	
where $\sigma = \exp[(2\pi L - \Delta V_A + \Delta V_B)/k_\text{B}T]$,
and $a$, $b$'s and $c$'s are dimensionless functions of load torque $L$ 
($a$, $c_0$, $c_A$ and $c_B$ depend also on $\kappa_A/D_0$ and $\kappa_B/D_0$).
Coefficients $a$, $b$'s and $c$'s are  given by expressions, 
too complicated to be presented here, 
containing  integrals of certain functions of $V_j(\theta) + L\theta$.
A few properties of these coefficients should be mentioned: 
(i) $b$'s and $c$'s are positive; 
(ii) $a = 0$ and $b_A = b_B$ for $L = 0$; 
(iii) $a > 0$ and $b_A > b_B$ for $L > 0$; 
and (iv) $a \to 0$, $b_A \to 1$, and $b_B \to 1$ in the limit of
large barrier of potential $V_0$. 
From the first two properties together with the definitions (\ref{eq:dbalance}) of 
$\rho_A$ and $\rho_B$, 
we observe that $\nu > 0$ and $r > 0$ for $\Delta\mu > 0$ 
in the absence of load torque ($L = 0$), 
as one would expect. 

It is emphasized that the dependences of $\nu$ and $r$ on $\rho_A$ and $\rho_B$ 
given by eqs.~(\ref{eq:v}) and (\ref{eq:r}) are quite 
\textit{general} in that
they are independent of potential profiles ($V_0$ and $V_1$), 
which affect only the values of coefficients 
($\sigma$, $a$, $b$'s and $c$'s). 
We also note that both $\nu$ and $r$ depend on
$\mu_A$ and $\mu_B$ separately, 
while in the two-state models for molecular motors 
worked out previously
\cite{lipowsky2000, lipowsky2003, lattanzi2001, parmeggiani99, 
astumian96} 
(and a single-state model \cite{sakaguchi2006} 
proposed recently) 
the chemical potentials come into play only through 
the difference $\Delta\mu = \mu_A - \mu_B$ 
(or $\Delta\mu = \mu_\text{ATP} - \mu_\text{ADP} - \mu_\text{Pi}$ if the reaction 
$\text{ATP} \rightleftarrows \text{ADP} + \text{Pi}$ 
is considered instead of $A \rightleftarrows B$).

In what follows we discuss the properties of the engine extracted from 
eqs.~(\ref{eq:v}) and (\ref{eq:r}) for a particular set of potentials 
$V_0$ and $V_1$. 
Here, for simplicity, we consider the piecewise linear functions shown 
in Fig.~\ref{fig:potentials}: 
the vertices of $V_0$ are located at $\theta = \theta_A$, $\theta_C$, and $\theta_B$ 
with $\theta_C$ satisfying $\theta_A < \theta_C < \theta_B$, 
and those of $V_1$ at $\theta = 0$, $\theta_A$, and $\theta_B$; 
the potential shapes are specified by parameters 
$\Phi_0 = V_0(\theta_C) - V_0(\theta_A)$, 
$W_0 = V_0(\theta_B) - V_0(\theta_A)$,
$\Phi_1 = V_1(0) - V_1(\theta_B)$, and 
$W_1 = V_1(\theta_A) - V_1(\theta_B)$.\cite{note}

\begin{figure}
	\begin{center}
	 \includegraphics[width = 6.5cm]{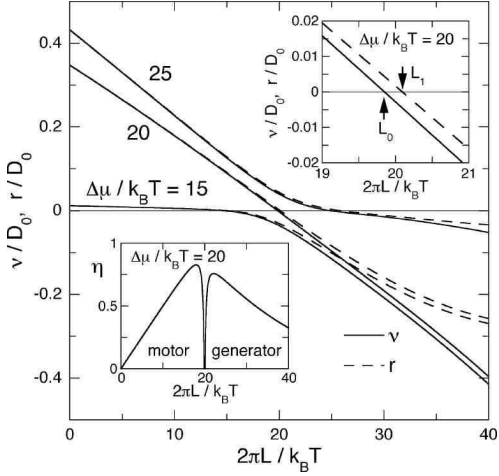}
	\end{center}
	\vspace{-2mm}
	\caption{
	The dependence of rotation rate $\nu$ (solid line) 
	and reaction rate $r$ (dashed line) on load torque $L$ 
	for	$\Delta \mu/k_\text{B}T = 15$, 20, and 25 with $\rho_B = 1.0$. 
	The parameters characterizing the potentials ($V_0$ and $V_1$) and the transition rates
	are chosen as follows: 
	$W_0 = W_1 = 10k_\text{B}T$, $\Phi_0 = 18k_\text{B}T$, $\Phi_1 = 40k_\text{B}T$, 
	$\theta_A = \pi/3$, $\theta_B = 4\pi/3$, $\theta_C = \pi$, 
	and $\kappa_A = \kappa_B = D_0$. 
	The upper inset is the magnification of a region where 
	$\nu \sim 0$ and $r \sim 0$ for $\Delta\mu/k_\text{B}T = 20$.
	The lower inset shows the efficiency $\eta$ for 
	$\Delta\mu/k_\text{B}T = 20$. 
	\label{fig:vreff}}
\end{figure}

It can be shown that  
eqs.~(\ref{eq:v}) and (\ref{eq:r}) are approximated by
\begin{equation}
	\nu \approx r 
	\approx D_0\rho_A\big/(c_0 + c_A\rho_A + c_{AB}\rho_A\rho_B)
	\label{eq:mmeq}
\end{equation}
in the absence of the load ($L = 0$) if 
$
	\exp\left(W_j/k_\text{B}T\right) \gg 1 \
$ 
($j = 0,1$), 
$
	\exp\left[(W_0 - \Phi_0)/k_\text{B}T\right] \gg 1
$,
$\kappa_A/D_0$ is not too small, 
and $\rho_B$ is not too large. 
The dependence on the concentration $\rho_A$ of fuel molecule 
for a fixed $\rho_B$ in this expression 
agrees with the one known as the Michaelis-Menten equation, 
and such a dependence of the rotation and reaction rates on the ATP concentration 
was observed for the F$_1$ motor \cite{yasuda2001, kinosita2004}. 
The dependence on $\rho_B$ predicted in eq.~(\ref{eq:mmeq}) may be
observed for the F$_1$ motor as the dependence on the ADP concentration.

Examples of the dependences of $\nu$ and $r$ on $L$ are shown in Fig.~\ref{fig:vreff}. 
Both $\nu$ and $r$ decrease monotonically with increasing $L$. 
The rotation rate $\nu$ becomes zero at 
a certain value $L_0$ of $L$, 
and the reaction rate $r$ becomes zero at 
a somewhat larger value $L_1$ 
(see the upper inset of Fig.~\ref{fig:vreff}).
Thus, the engine works as a motor for $0 \le L < L_0$ 
and as a generator for $L > L_1$, 
whereas it wastes both the chemical and mechanical energies 
for $L_0 < L < L_1$. 
It is not difficult to see that 
$L_0 < \Delta\mu/2\pi < L_1$ 
from property (iii) mentioned above, 
and that both $L_0$ and $L_1$ 
tends to $\Delta \mu/2\pi$ (therefore the useless interval of $L$ 
vanishes) 
in the tight coupling limit 
[$\exp(-\Phi_0/k_\text{B}T) \to 0$] according to property (iv). 

In the examples shown in Fig.~\ref{fig:vreff}, 
the mechanochemical coupling is almost tight ($\nu \simeq r$) for small load. 
However, the difference between $\nu$ and $r$ becomes apparent 
(the coupling becomes looser) 
as $L$ is increased. 
This is because the effective barrier height $\Phi_0 - W_0 - L(\theta_B - \theta_C)$ 
of potential $V_0$ seen from the location $\theta = \theta_B$ decreases 
and hence the leftward dashed passes in Fig.~\ref{fig:potentials} 
are taken more frequently for larger $L$.

The \textit{efficiency} $\eta$ of energy transduction  
is defined as $\eta = 2\pi\nu L/r\Delta\mu$ 
for motor and $\eta = r\Delta\mu/2\pi\nu L$ for generator 
\cite{parmeggiani99}. 
These expressions may be written as
\begin{equation}
\label{eq:efficiency}
	\eta = \chi\eta_0,
\end{equation}
where $\chi = \nu/r$ ($r/\nu$) is 
the ``tightness'' of mechanochemical coupling,  
and $\eta_0 = 2\pi L/\Delta\mu$ ($\Delta\mu/2\pi L$) is 
the efficiency in the tight coupling limit 
for motor (generator). 
Note that larger $\chi$ indicates tighter mechanochemical 
coupling, and we have $\chi = 1$ in the tight coupling limit.

\begin{figure}
	\begin{center}
		\includegraphics[width = 5cm]{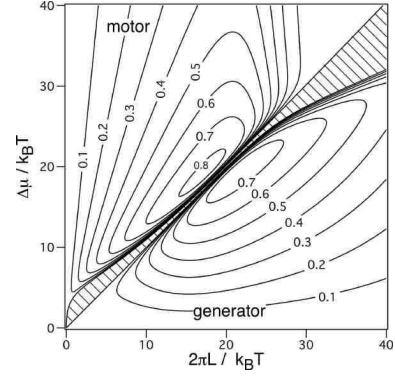}
	\end{center}
	\vspace{-2mm}
	\caption{
	The dependence of the efficiencies of motor (the upper left portion)
	and of generator (the lower right portion)
	on $L$ and $\Delta\mu$ is shown as a contour plot for $\rho_B = 1$. 
	The other parameters are the same as the ones in Fig.~\ref{fig:vreff}. 
	In the hatched region the engine works as neither a motor nor a generator 
	(it is useless). 
	\label{fig:contour}}
\end{figure}

An example of $\eta$ as a function of $L$ is depicted in the 
lower inset of Fig.~\ref{fig:vreff}, where we observe that 
$\eta$ for motor (generator) has a maximum 
near the ``stall'' load $L_0$ 
($L_1$). 
In the tight coupling limit, 
the maximum efficiency of $\eta_0 = 1$ is achieved 
at $L = \Delta \mu/2\pi \mp 0$ (the minus sign for motor 
and the plus sign for generator). 
In the case of loose coupling, 
the maximum value of $\eta$ tends to be larger for 
larger $\chi$. 
Since the tightness decreases  
with increasing $L$ as explained above, 
the maximum of $\eta$ for motor is larger than that for generator in this example. 

The dependence of $\eta$ on $L$ and $\Delta\mu$ for a particular choice of $\rho_B$ 
is shown as a contour plot in Fig.~\ref{fig:contour}. 
In this example, the largest efficiencies of motor and generator 
are achieved in the condition $\Delta\mu \sim 2\pi L \sim 20k_\text{B}T$ far from equilibrium ($\Delta\mu =  L = 0$); 
a similar observation was made for related 
models by Parmeggiani \textit{et al} \cite{parmeggiani99}. 
Note that, in the tight coupling limit, we have the maximum 
value of $\eta_0 = 1$ on the diagonal line 
$\Delta\mu = 2\pi L$.  

\begin{figure}
	\begin{center}
	 	\includegraphics[width = 6.5 cm]{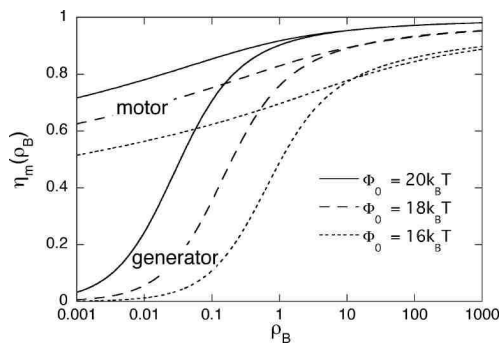}
	\end{center}
	\vspace{-2mm}
	\caption{
	The maximum efficiencies $\eta_\text{m}(\rho_B)$ 
	of motor and of generator for given $\rho_B$ are 
	plotted against $\rho_B$ for different choices of $\Phi_0$. 
	The other parameters are the same as the ones in Fig.~\ref{fig:vreff}.
	\label{fig:emax}}
\end{figure}

We have obtained qualitatively similar patterns of contour lines to the one shown in Fig.~\ref{fig:contour} for other choices of $\rho_B$, 
although the locations and the hights of the peaks change as $\rho_B$ is varied.
Let $\eta_\text{m}(\rho_B)$ be the maximum value of $\eta$ obtained 
by adjusting $L$ and $\Delta\mu$ for a given $\rho_B$. 
Figure~\ref{fig:emax} shows $\eta_\text{m}$ as a function of $\rho_B$ 
for three choices of the barrier height $\Phi_0$ of potential $V_0$. It is noted that $\eta_\text{m}$ increases with $\rho_B$ 
for both motor and generator, 
and saturates to a certain value---%
the upper limit of efficiency achieved by the engine 
(characterize by $V_0(\theta)$, $V_1(\theta)$, 
$\kappa_A$, $\kappa_B$ and $D_0$).

The dependence of $\eta_\text{m}(\rho_B)$ on $\rho_B$ shown 
in Fig.~\ref{fig:emax} may be understood qualitatively 
as follows. 
Remember that larger $\eta_\text{m}$ is expected 
for larger tightness $\chi$ of mechanochemical coupling. 
As the the concentration $\rho_B$ of molecule~$B$ 
is increased, 
the binding of molecule $B$ (transition from state~0 
to state~1 at $\theta = \theta_B$) 
occurs more frequently, 
which in effect leads to the decrease in 
the chance of taking the leftward dashed passes 
in Fig.~\ref{fig:potentials}.  
Therefore, the tightness $\chi$ and hence $\eta_\text{m}$ 
will increase with $\rho_B$, which is consistent with 
what we see in Fig.~\ref{fig:emax}.  
In the case of motor,  
the dissociation of molecule $B$ occurs more frequently than 
the binding, and the former is not affected by $\rho_B$, 
while the binding of molecule $B$ is essential for generator. 
This explains the stronger dependence of $\eta_\text{m}$ on $\rho_B$ for generator than for motor observed 
in Fig.~\ref{fig:emax}.

It may be worth remarking that the present model may be 
viewed as a motor driven by ion-flow across a membrane:
\cite{xing2004, xing2006} 
the transition at $\theta_A$ can be viewed as ion 
exchange with the outside of the membrane, 
and the one at $\theta_B$ as ion exchange with the inside; 
in this case only one chemical species is involved.

In summary, we have proposed a minimal model for 
molecular chemical engines that properly takes into account 
the effects of fuel and waste molecules.
The model is simple enough to work out various properties
of the engine such as the efficiency of energy transduction. 
The detailed analysis of the model, 
including the derivations of various results presented here, and its extensions to situations other than the rotary motor 
will be reported in future publications.

We would like to thank A.~Parmeggiani, J.~Prost, J.-F.~Joanny, K.~Sekimoto, E.~Muneyuki, H.~Noji, 
H.~Higuchi, F.~Matsubara, and M.~Sasaki for useful discussions and comments. 
This work was supported in part by
the Grants-in-Aid for
Scientific Research in Priority Areas from the Japan Ministry
of Education, Culture, Sports, Science and Technology.


\begin{thebibliography}{99}
\bibitem{note}
Note that 
difference $\Delta V(\theta) \equiv V_1(\theta) - V_0(\theta)$ 
is not important---%
because it depends on an arbitrarily chosen (common) zero of 
$\mu_A$ and $\mu_B$ (adding a constant $\mu_0$ to $\mu_A$ 
and $\mu_B$ amounts to increase in $\Delta V(\theta)$ by $\mu_0$),  
while the transition rates depends on $\mu_\alpha$ and $V_j$ 
only through $\mu_\alpha - \Delta V(\theta_\alpha)$, 
which are not affected by such a choice.
\end{thebibliography}
\end{document}